\documentclass[amssymb,aps,floatfix,floats,preprintnumbers,showpacs,twocolumn,prl,reprint]{revtex4-1}
\usepackage{rcs}
\usepackage{graphicx}
\usepackage{dcolumn}
\usepackage{bm}
\usepackage[applemac]{inputenc} 
\begin{document}
\newcommand{\bsig}{{\bm \sigma}}
\newcommand{\btau}{{\bm \tau}}
\newcommand{\bmu}{{\bm \mu}}
\newcommand{\bs}{{\bm s}}
\newcommand{\bS}{{\bm S}}
\newcommand{\bQ}{{\bf Q}}
\newcommand{\bk}{{\bf k}}
\newcommand{\bq}{{\bf q}}
\newcommand{\bM}{{\bf M}}
\newcommand{\bR}{{\bf R}}
\newcommand{\bt}{{\bf t}}
\newcommand{\bars}{\bar{s}}
\newcommand{\la}{\langle}
\newcommand{\ra}{\rangle}
\newcommand{\da}{\dagger}
\newcommand{\up}{\uparrow}
\newcommand{\down}{\downarrow}
\newcommand{\lc}{\lowercase}
\newcommand{\no}{\nonumber}
\newcommand{\be}{\begin{equation}} 
\newcommand{\ee}{\end{equation}}
\newcommand{\bea}{\begin{eqnarray}} 
\newcommand{\eea}{\end{eqnarray}}
\newcommand{\Q}{($0,0,\frac{2\pi}{c}$)}
\newcommand{\mul}{$\bmu_{loc}$}
\newcommand{\PPS}{Pr$_{3}$Pd$_{20}$Si$_{6}$}
\def\bra#1{{\langle #1 \vert}}
\def\ket#1{{\vert #1 \rangle}}
\def\x#1{{\sigma_{#1}^{x}}}
\def\z#1{{\sigma_{#1}^{z}}}
\def\y#1{{\sigma_{#1}^{y}}}
\title{The role of hyperfine coupling in magnetic and quadrupolar ordering of \PPS}
\author{L. Steinke$^1$, K. Mitsumoto$^{1,2}$, C.~F.\ Miclea$^{1,3}$, F. Weickert$^1$, A. D\"onni$^4$, M. Akatsu$^{2,5}$, Y. Nemoto$^2$, T. Goto$^2$, H. Kitazawa$^4$, P. Thalmeier$^1$, and M. Brando$^1$}
\affiliation{$^1$Max Planck Institute for Chemical Physics of Solids, D-01187 Dresden, Germany \\
$^2$Graduate School of Science and Technology, Niigata University, Niigata 950-2181, Japan \\
$^3$National Institute for Materials Physics, 077125 Bucharest-Magurele, Romania \\
$^4$National Institute for Materials Science, Tsukuba 305-0047, Japan \\
$^5$Dresden High Magnetic Field Laboratory, Helmholtz- Zentrum Dresden-Rossendorf, D-01314 Dresden, Germany} 
\date{\today}
\begin{abstract}
We study the ternary clathrate \PPS\,in specific heat and AC-susceptibility measurements on a high-quality single crystal, distinguishing antiferromagnetic (AFM) and antiferroquadrupolar (AFQ) ordering on two sublattices of inequivalent Pr sites. The specific heat shows the direct involvement of nuclear spin degrees of freedom in the AFM ordering, which is well supported by our calculation of the hyperfine level scheme without adjustable parameters.
\PPS\,is therefore one of the rare materials where the nuclear moments are involved in the formation of the magnetic ground state.
\end{abstract}
\maketitle
\section{Introduction}
The ground state of solids is commonly described without referring explicitly to the nuclear degrees of freedom. In lanthanide compounds there are, however, rare examples where nuclear moments directly influence the magnetic ground states. This may happen when the dipole hyperfine (HF) coupling $A$ of nuclear spin $\boldsymbol{I}$ and total angular momentum $\boldsymbol{J}$ of 4$f$ electrons is sufficiently large. In Pr compounds, where $A$ has the exceptionally high value of 52  mK \cite{Kondo:1961,Bleaney:1963}, the HF interaction becomes even decisive for the magnetic order of the ground state, if the crystalline electric field (CEF) effect favors a 4$f$ singlet. For example, in pure Pr metal incommensurate induced antiferromagnetism requires the assistance of the large HF coupling \cite{Moller:1982,Eriksen:1983}. HF interactions also play an essential role in the singlet ground state system PrCu$_2$\cite{Andres:1972}, and they were found to induce magnetic ordering of a $S =1/2$ quantum magnet on a Kagom\'e lattice~\cite{Karaki:2010}. Importantly, the magnetic transition temperatures of such coupled systems are usually on the order of a few milli-Kelvin, below the base temperature achievable in most laboratories. Here we provide evidence that in \PPS\,not just the electronic but also the nuclear degrees of freedom are involved in a magnetic transition at $T_{N1} = 0.14$\,K, a temperature that is easily reached by a standard dilution refrigerator. This material is therefore an excellent candidate to investigate the role of HF coupling in the formation of magnetic ground states.   
\begin{figure}
\includegraphics[width=\columnwidth]{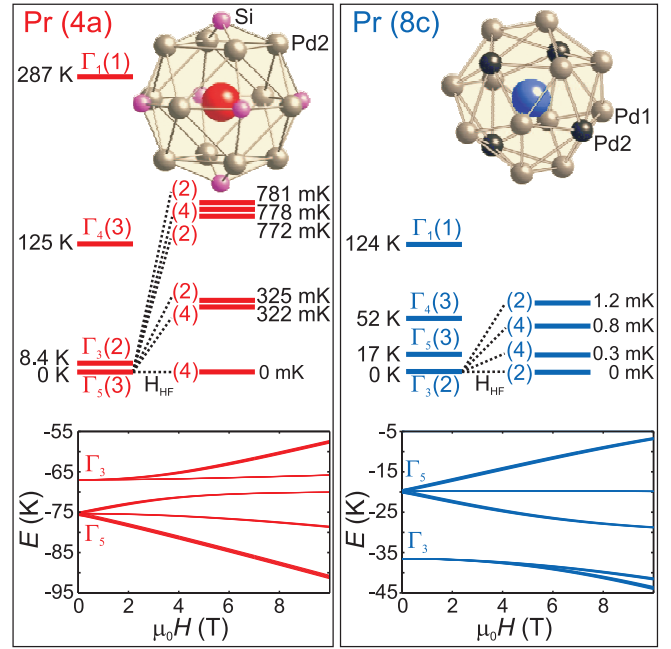}
\caption{(Color online) Cage structure and HF level scheme for Pr 4a ($O_h$ symmetry) and 8c sites ($T_d$ symmetry). The energy levels are calculated at $H = 0$, using the CEF Hamiltonian obtained from INS~\cite{Doenni:2011}. Inset graphs: calculated $H$-dependence of the lowest two CEF levels for $H\,||\,[110]$. The line thickness indicates the scale of HF splitting.}
\label{Fig1}
\end{figure}

Furthermore, \PPS\,belongs to a new class of ternary clathrates R$_3$Pd$_{20}$X$_6$ (R = rare earth, X = Si,Ge) that has attracted considerable interest after the discoveries of quadrupolar (QP) ordering~\cite{Kitagawa:1996} and thermally activated rattling~\cite{Nemoto:2003} of Ce ions in Ce$_3$Pd$_{20}$Ge$_6$. R atoms in these cage compounds reside on two sublattices of inequivalent sites: a face centered cubic lattice of 4a sites with cages of cubic $O_h$ symmetry (Fig.~\ref{Fig1}, left), and a simple cubic lattice of 8c sites, where the cages have cubic $T_d$ symmetry (Fig.~\ref{Fig1}, right). The occupancy ratio of 8c:4a is 2:1. Depending on the 8c and 4a CEF ground states, a rich variety of multipolar orders may appear at low temperatures, as witnessed by measurements of the elastic constants~\cite{Kobayashi:2008,Akatsu:2009,Ano:2012}, as well as thermodynamic and transport studies~\cite{Custers:2012}.
The \PPS\,clathrate investigated here is particularly interesting, since the CEF ground state entails the possibility of both magnetic and QP order.
\begin{figure}
\includegraphics[width=\columnwidth]{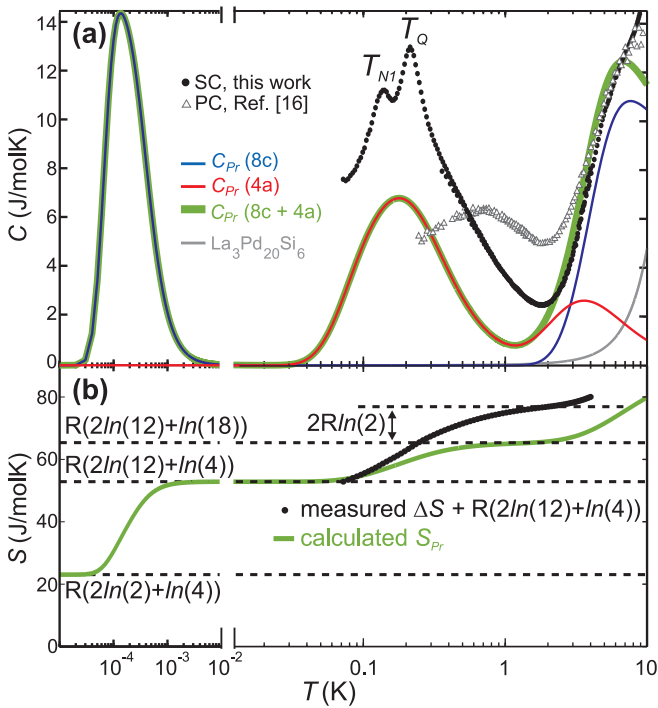}
\caption{(Color online) \textbf{(a)}: Specific heat $C(T)$ of \PPS\,at \mbox{$B = 0$\,T}, measured for SC (black symbols) and PC samples (gray symbols). While the SC measurements clearly resolve two peaks at $T_Q=0.21$\,K and $T_{N1}=0.14$ K, these features are absent in the PC data. Instead, a broad peak is observed around 0.6\,K. Solid lines show the calculated specific heat for the 4a (red) and 8c sites (blue), and the sum of the two (green). Non-magnetic contributions (gray line) are estimated from $\mathrm{La_3Pd_{20}Si_6}$~\cite{Kitagawa:1997}.  \textbf{(b)}: Calculated entropy $S$ (green line), compared to the entropy obtained from the measured $C/T$ (black). The additional entropy release of $2R\ln{2}$ between 0.07 and 2\,K indicates fully developed ordering of the 8c sites, lifting the two-fold degeneracy of the $\Gamma_3$ ground state.}
\label{Fig2}
\end{figure}
Under the cubic symmetry of the CEF Hamiltonian $\mathcal{H}_{CEF}$, the 9-fold Pr 4$f$ multiplet ($J = 4$) splits into two triplets ($\Gamma_4$ and $\Gamma_5$), one $\Gamma_3$ doublet and one $\Gamma_1$ singlet. Inelastic neutron scattering (INS)~\cite{Doenni:2011} has shown different CEF ground states for the two Pr sites: a $\Gamma_5$ triplet on the 4a sites and a nonmagnetic $\Gamma_3$ doublet on the 8c sites, with electric quadrupoles as the lowest-order moments. In this Letter, we study both the magnetic and quadrupolar ordering of the two Pr sublattices in $\mathrm{Pr_3Pd_{20}Si_6}$, and we show how the large on-site HF coupling between the 4$f$ and the $^{141}\mathrm{Pr}$ nuclear moments influences these ordering transitions.
\section{Single-ion calculations}
We calculate the HF level scheme, as described in Ref.~\onlinecite{Aoki:2011}, by adding the on-site HF dipole interaction $A\boldsymbol{I}\cdot\boldsymbol{J}$ between the 4$f$ total angular momentum and the $^{141}\mathrm{Pr}$ nuclear spin ($I = 5/2$) to the experimental CEF Hamiltonian from Ref.~\onlinecite{Doenni:2011}:
\begin{equation}
\mathcal{H} = \mathcal{H}_{CEF} + A\boldsymbol{I}\cdot\boldsymbol{J}+(g_J\mu_B\boldsymbol{J}-g_N\mu_N\boldsymbol{I})\cdot \mu_0\boldsymbol{H}.
\label{H}
\end{equation}
An external magnetic field $\mu_0\boldsymbol{H}$ induces electronic and nuclear Zeeman splitting $(g_J\mu_B\boldsymbol{J}- g_N\mu_N\boldsymbol{I})\cdot \mu_0\boldsymbol{H}$, where $g_J=4/5$, $g_N=1.72$, and $\mu_B$ or $\mu_N$ represents the Bohr or nuclear magneton, respectively. Using the coupling constant $A = 52$ mK from Refs. \cite{Kondo:1961, Bleaney:1963}, we calculate the eigenenergies  of the Hamiltonian (\ref{H}) shown in Fig.~\ref{Fig1}. For the 4a site, where the CEF ground state is a $\Gamma_5$ triplet with a nonzero magnetic dipole moment, the large on-site HF dipole interaction significantly affects the zero field energy spectrum and causes a reconstruction of the $\Gamma_5$ triplet into a quartet at 0 K, a sextet at 0.323\,K, and an octet at 0.778\,K, with additional splitting on the order of mK. The non-magnetic $\Gamma_3$ doublet on the 8c-site shows only small HF splitting of 4 - 5 mK due to second order effects in $A$. The magnetic field dependence of the lowest energy levels is dominated by the electronic Zeeman splitting of the $\Gamma_5$ triplets. Large fields applied along the $[110]$-direction lead to a splitting of the non-Kramers $\Gamma_3$ doublets above $\mu_0H \approx 3$ T for the 4a sites and $\approx 6$ T for the 8c sites.
\section{Experimental results}
\begin{figure}
\includegraphics[width=\columnwidth]{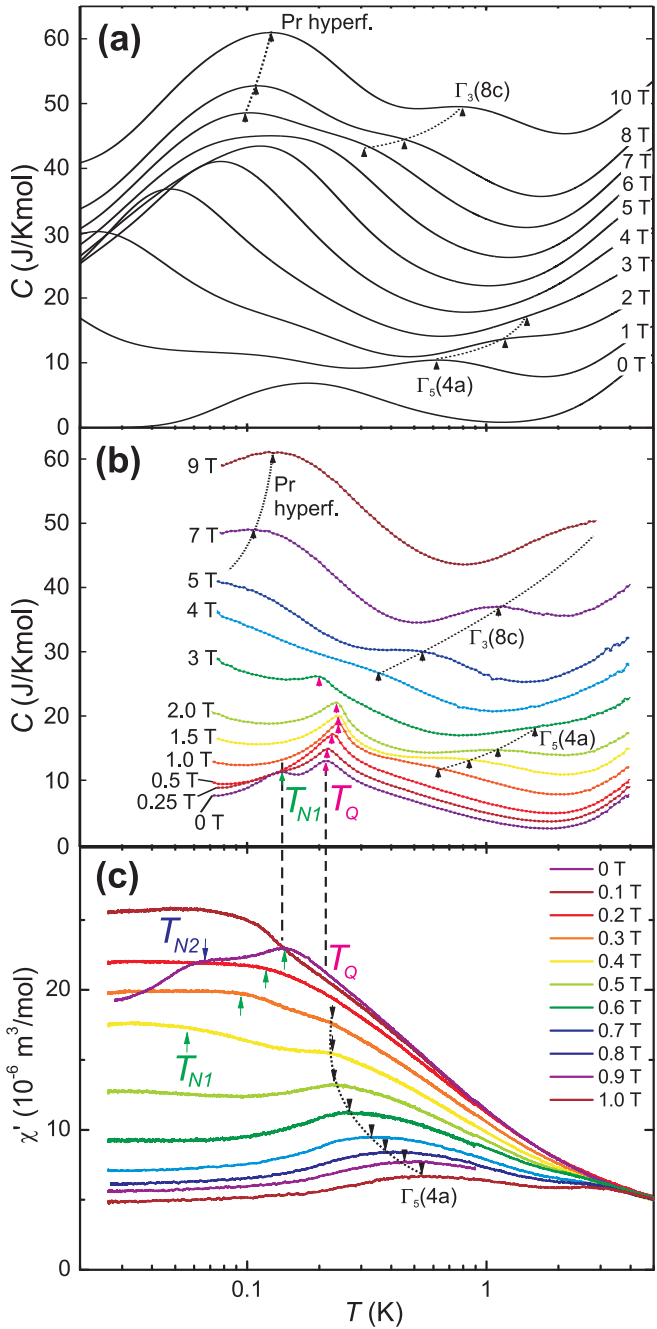}
\caption{(Color online) \textbf{(a)} Calculations of $C_{Pr} = C_{Pr}(4a)+C_{Pr}(8c)$ for $H\,||\,[110]$. \textbf{(b)}$T$-dependent specific heat $C$, measured at various fields $0\,\mathrm{T}\leq\mu_0H\leq9\,\mathrm{T}$. For clarity, the curves in panels \textbf{(a)} and \textbf{(b)} are shifted by $+4$\,J/Kmol per T relative to $H=0$. $T_{N1}$ and $T_Q$ indicate the critical temperatures for phase transitions at low fields. Comparing panels \textbf{(a)} and \textbf{(b)}, further peaks in $C(T)$ and $\chi'(T)$ at higher fields are identified as Schottky peaks from Zeeman splitting of the electronic and HF ground state. \textbf{(c)} AC-susceptibility $\chi'(T)$ at dc-fields between 0 and 1.2\,T. The peak at $T_{N1}$ appears in $\chi'(T)$ while the peak at $T_Q$ is absent. An additional transition is found at $T_{N2} = 0.06$\,K.}
\label{Fig3}
\end{figure}
To distinguish magnetic and non-magnetic phases in \PPS, and to obtain a complete $H-T$-phase diagram at low temperatures, we combined measurements of the specific heat $C$ and magnetic AC-susceptibility $\chi_{AC}$.
All measurements were made on the same $\mathrm{Pr_3Pd_{20}Si_6}$ single crystal (SC), which was fabricated by floating zone melting using an image furnace with two ellipsoidal mirrors. The specific heat was measured with a semiadiabatic compensated heat-pulse method \cite{Wilhelm:2004}, using a $^3\mathrm{He}/^4\mathrm{He}$ dilution refrigerator between 0.07 and 4\,K, and a Quantum Design PPMS equipped with a $^3\mathrm{He}$-insert between 0.3 and 10\,K.
The $\chi_{AC}$ measurements between 0.02 and 4\,K were carried out in a Kelvinox 400 $\mathrm{^3He/^4He}$-dilution refrigerator, with a modulation field of 15\,$\mu$T at 113.7\,Hz. Absolute values of $\chi_{AC}$ were determined by matching the data above 2 K to calibration measurements in a Quantum Design SQUID VSM.

Fig.~\ref{Fig2} shows the specific heat $C$ measured at zero magnetic field. We observe two distinct peaks at $T_Q=0.21$ K and $T_{N1}=0.14$ K, consistent with phase transitions found in previous ultrasound measurements of $\mathrm{Pr_3Pd_{20}Si_6}$ SC~\cite{Kobayashi:2008,Akatsu:2009}. These transitions were not observed in specific heat data of polycrystalline (PC) samples, prepared by arc-melting and annealing~\cite{Kitagawa:1997}. The 3-20-6 phase is known to possess a very narrow homogeneity range~\cite{Kitagawa:1999}, causing a strong sample dependence of physical properties. Our specific heat data show that high-quality SC are necessary to study the low-temperature phases in $\mathrm{Pr_3Pd_{20}Si_6}$. To quantify the specific heat related to the ordering of both Pr sites, all non-magnetic contributions must be subtracted. These are estimated from reference measurements of the isostructural non-magnetic compound $\mathrm{La_3Pd_{20}Si_6}$~\cite{Kitagawa:1997} (gray line in Fig.~\ref{Fig2}), and turn out to be negligible in the region of interest $T \leq 2$ K. Next, we compare the measured specific heat to the single-ion specific heat $C_{Pr}$ (Fig.~\ref{Fig2}, green line) calculated for the two Pr sites using the HF level schemes in Fig.~\ref{Fig1}.
\begin{figure}
\includegraphics[width=\columnwidth]{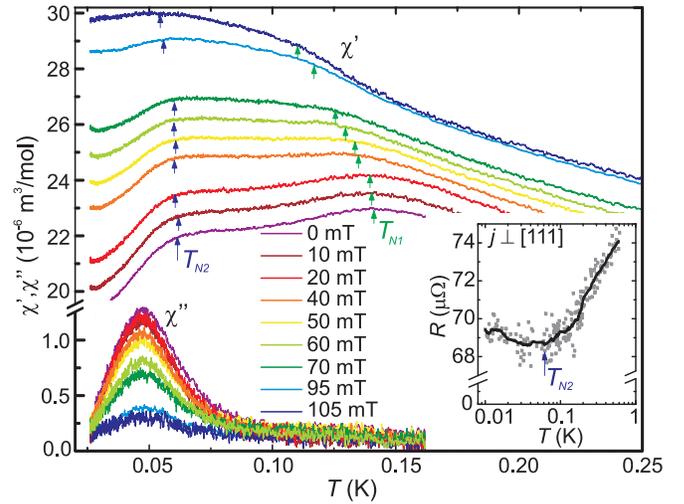}
\caption{(Color online) Real part $\chi'(T)$ and imaginary part $\chi''(T)$ of the magnetic AC-susceptibility, at various dc fields $0\leq\mu_0H\leq105$\,mT. $\chi'(T)$ curves are offset by $+0.4\times 10^{-6}\,\mathrm{m^3/mol}$ per $10$\,mT relative to $H = 0$. Green and blue arrows indicate the transitions at $T_{N1}$ and $T_{N2}$, respectively. Inset: four-point resistance measurement of the same crystal at 0\,T, showing a change of slope around $T_{N2}$.}
\label{Fig4}
\end{figure}
We find that the low-temperature specific heat is greatly enhanced by excitation of nuclear degrees of freedom. Integrating the measured $C/T$ over $T$ (Fig.~\ref{Fig2}, bottom panel), we obtain $\Delta S = 24.05\,\mathrm{J/Kmol}$ for the entropy between 0.07 and 2\,K. The degeneracy of the $\Gamma_5$ and $\Gamma_3$ CEF ground states at the 4a and 8c sites can only account for $R\ln3+2R\ln2 = 20.66$ J/Kmol, if both sites order completely. Without the involvement of nuclear degrees of freedom, the measured $\Delta S$ would therefore be in strong contrast to the CEF ground state obtained from INS \cite{Doenni:2011}.
However, our calculation shown in Fig.~\ref{Fig2} predicts a large specific heat peak due to the thermalization of HF multiplets on the 4a sites, which coincides with the low-$T$ phase transitions. The expected entropy release from the thermal occupation of the 4a site HF sextet at 0.323\,K and the octet at 0.778\,K is $R(\ln(18)-\ln(4))=12.51$ J/Kmol. Subtracting this value from the measured $\Delta S$ leaves an entropy of 11.54 J/Kmol, which is very close to $R\ln(4)=2R\ln(2)=11.53$\,J/Kmol.

The additional entropy release could either be attributed to ordering of the HF ground state quartet of the 4a sites ($\Delta S = R\ln(4)$), or to fully developed ordering of the doubly occupied 8c sites, lifting the two-fold degeneracy of the $\Gamma_3$ ground state($\Delta S = 2R\ln(2)$). $\chi_{AC}$-measurements can distinguish these magnetic or non-magnetic types of ordering. Fig.~\ref{Fig3} compares $T$-dependent specific heat $C(T)$ (panel b) and the real part $\chi'$ of $\chi_{AC}$ (panel c) to calculations of Schottky anomalies in $C(T)$ (panel a), based on the previously calculated single-ion level scheme (cf. Fig.~\ref{Fig1}).
The peak positions in $C(T)$ agree well with the calculations. Only the field-induced splitting of the $\Gamma_3$ doublets is significantly underestimated by the numerical results, where the measurement at 7\,T closely resembles the calculated curve at 10\,T. 
We observe a common peak in $C$ and $\chi'$ at $T_{N1} = 0.14$ \,K, whereas the second peak at $T_Q = 0.21$\,K only appears in the specific heat for all the applied magnetic fields.
This implies ordering of the non-magnetic 8c sites at $T_Q$, where the lowest-order moments are electric quadrupoles, and attributes the entropy release $\Delta S = 2R\ln(2)$ to fully developed QP order on the 8c sublattice. While the $\Gamma_5$ ground state of the 4a sites would in principle support non-magnetic QP order as well, the transition would be revealed by the susceptibility at finite fields, due to a field-induced mixing of dipolar and QP moments~\cite{Kiss:2003, Pourret:2006, Custers:2012}. Therefore, the non-magnetic transition at $T_Q$ is interpreted as QP ordering of the 8c sites, and the magnetic transition at $T_{N1}$ as AFM ordering of the 4a sites. The HF ground state of the 4a sites remains effectively four-fold degenerate down to 0.07\,K.

In addition to the transitions at $T_Q$ and $T_{N1}$, we observe a third feature at $T_{N2} = 0.06$\,K, presented in more detail in Fig.~\ref{Fig4}. A drop of $\chi'$ below $T_{N2} = 0.06$\,K is accompanied by a peak of the imaginary susceptibility $\chi''$, which indicates dissipative processes. With increasing $H$, $T_{N2}$ shifts slightly to lower temperatures and the anomalies in both $\chi'(T)$ and $\chi''(T)$ gradually disappear. A four-point resistance measurement (Fig.~\ref{Fig4}, inset) shows weak $T$-dependence with a slight increase of $\approx 5\,\mu\Omega$ between 0.1 and 0.02\,K, and a change of slope occurs around $T_{N2}$. Without further measurements it is difficult to establish the nature of this transition. 

\begin{figure}
\includegraphics[width=\columnwidth]{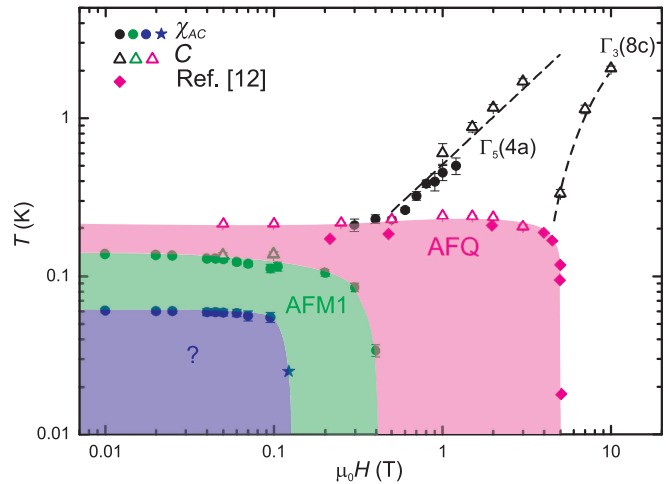}
\caption{(Color online) Phase diagram of \PPS\,derived from specific heat (triangles) and AC-susceptibility $\chi_{AC}(T)$ (circles) and $\chi_{AC}(H)$ (star symbol). Results from Ref.\cite{Akatsu:2009} (diamonds) are added for comparison. We identify an AFQ ordered phase (magenta) of the Pr 8c sites, followed by AFM ordering of the 4a sites (green) below 0.14\,K. The additional phase below $T_{N2} = 0.06$\,K (blue) is yet unknown.}
\label{Fig5}
\end{figure}
Fig.~\ref{Fig5} shows the phase diagram derived from the $T$- and $H$-dependence of features in $C$ and $\chi'$. The highest temperature phase boundary agrees very well with previous results from measurements of the elastic constants~\cite{Akatsu:2009}. From the shape of the phase boundary and symmetry arguments, this transition was interpreted as AFQ ordering of $O_2^0=\frac{1}{2}(2J_z^2-J_x^2-J_y^2)$ moments at the 8c sites, where the $\Gamma_3$ ground state does not support a magnetic dipolar order parameter \cite{Akatsu:2009}. Our observations of a non-magnetic transition and an entropy release consistent with full ordering of the 8c sublattice confirm this interpretation. The phase below $T_{N1} = 0.14$\,K is attributed to AFM ordering of the 4a sites. The newly discovered third transition below $T_{N2} = 0.06$\,K must involve magnetic moments of the 4a sites as well, since the transition shows up in the zero field susceptibility. A first-order incommensurate-to-commensurate AFM transition is one possible scenario, which would be compatible with both the finite $\chi''$ and the change of slope in the electrical resistance, showing only weak $T$-dependence below $T_{N2}$. The ordering could also involve the effective magnetic moment of the 4a-site ground state quartet (pseudospin 3/2 \cite{Aoki:2011}), which is still degenerate above $T_{N2} = 0.06$\,K.
\section{Conclusions}
We have shown how specific heat and AC-susceptibility measurements can be effectively combined to study magnetic and non-magnetic multipolar phases in Pr$_3$Pd$_{20}$Si$_6$. We confirmed previously reported AFQ and AFM ordering and discovered an additional magnetic phase transition at 0.06\,K. 
Our results further demonstrate that \PPS\,is another one of the rare examples where nuclear degrees of freedom are directly involved in the ordering of electronic 4$f$ multipoles. Since the calculated HF peak in the specific heat of the 4a sites coincides exactly with the experimentally observed AFM phase transition, it is likely that the HF interaction plays an essential role in the AFM order.
The present theoretical analysis of specific heat and entropy is limited to single ion CEF potentials and HF interactions. A more realistic description should include inter-site dipolar and quadrupolar interactions and their corresponding molecular fields \cite{takimoto:08}. This will give a deeper insight on how the large Pr HF coupling influences the thermodynamics of  magnetic and quadrupolar order in the Pr clathrate.
\section{Acknowledgements}
We would like to thank C. Geibel, T. Herrmannsd\"orfer, J. Kitagawa, M. Rotter and A. Steppke for valuable discussions, as well as T. Lühmann, M. Nicklas and C. Klausnitzer for technical support.
CFM acknowledges support from  PN-II-ID-PCE-2011-3-1028.

\end{document}